# The reltation between the electronic structure and thermoelectric transport properties for Zintl compounds $M_2Zn_5As_4$ (M=K, Rb)

Gui Yang[*](1)(2), jueming Yang, Yuli Yan, Yuanxu Wang (1) [*]

(1) *Institute for Computational Materials Science, School of Physics and Electronics, Henan University, Kaifeng 475004, China.*

(2) *College of physics & Electrical Engineering, Anyang Normal University, Anyang Henan,455000 China*

**Abstract:** The electronic structure and thermoelectric properties of $M_2Zn_5As_4$ (M=K,Rb) are studied by the first principles and the semiclassical BoltzTrap theory. The results calculated by different exchange-correlation potentials show that these compounds are semiconductors with the indirect band gap. The combination of heavy and light bands near the valence band maximum is conducive to improve the thermopower because the strong dispersive band (light bands) favors electrical conductivity $\sigma$, and the weakly dispersive band (heavy bands) sets a low-energy scale that can enhance the Seeback coefficient *S*. In deed, our calculations show that relatively large Seeback coefficients, high conductivities, and the big " maximum" thermoelectric figures of merit $Z_e T = \dfrac{S^2 \sigma T}{\kappa_e}$ exist for $Rb_2Zn_5As_4$ in the low hole-doping region. Together with the expected low thermal conductivity, these compounds are suggested to be a new good thermoelectric materials especially by hole-doped for the potential applications.

**Keywords:** Zintl-phase compounds; thermoelectric properties;

*Email: wangyx@henu.edu.cn

# 1. Introduction

More than half the energy consumed nationwide is lost as heat, the recovery of even a small fraction would significantly impact global energy use. The direct energy conversion between heat and electricity is attractive for many application in power generation and heat pumping[1-8]. This process is based on the "Seeback effect", which is the appearance of an electrical voltage caused by a temperature gradient ($\Delta T$) across a material. Like all heat engines, the power generation efficiency ($\eta$) is thermodynamically limited by the Carnot efficiency ($\Delta T / T_{hot}$)[9].

$$\eta = \frac{\Delta T}{T_{hot}} \frac{\sqrt{1+ZT_{avg}}-1}{\sqrt{1+ZT_{avg}}+\frac{T_{cold}}{T_{hot}}} \tag{1}$$

Where $T_{hot}$ and $T_{cold}$ are the temperature of the hot and cold ends in a thermoelectric device. The term $\sqrt{1+ZT_{avg}}$ varies with the average temperature $T_{avg}$. Equation (1) shows that the device efficiency generally relies on the material figure of merit, $ZT = \frac{S^2 \sigma T}{\kappa}$, the quantity $S^2\sigma$ is called the thermopower. The request to get a large ZT material needs to have a large Seeback cofficient ($S$), high electrical conductivity($\sigma$) and low thermal conductivity $\kappa = \kappa_{el} + \kappa_{l}$, simultaneously. $\kappa_{el}$ and $\kappa_{l}$ respectively indicate the electron and lattice components of the thermal conductivity. However, it is difficult to balance because the electrical conductivity and the electronic part of the thermal conductivity are generally connceted by the Wiedemann-Franz law.

Zintl-phase compounds appear to be promising candidates for thermoelectric applications. These materials provide a good balance between the thermoepower and thermal conductivity because they combine the favorable features of chemical tunability and the simple electron counting rules associated with valence compounds. These unusual materials were first characterized by Eduard Zintl in the 1930s. Typically, the containing arsenic ternary Zintl-phase compounds[10-12] are of the form $A_m B_x As_y = (A^{n+})_m B_x^{(\frac{3y-mn}{x})+} As_y^{3-}$, which can also rewrite as $A_m[B_x As_y]$, where the atomic $A$ (alkali, alkali-earth or rare earth), and $B$ (transitional metal) are electropositive cations.

The cation *A* donates its valence electrons to the cluster $[B_xAs_y]$ which in turn form the coordinated structure capable of accommodating extra electrons in its bonds. Therefore, the Zintl phase shows the coexist characteristic of both ionic and covalent to the bonding picture. Such a complex crystal structure enchances the phonon scattering and consequently result in a low lattice thermal conductivity. In addition, the rich solid-state chemistry of Zintl phases enables numerous possibilities for chemical substitutions and structural modifications that allow the fundamental transport parameters (carrier concentration, mobility, effective mass, and lattice thermal conductivity) to be modified for enchanced thermoelectric performance.

Lots of Zintl phases and related compounds have been proven to be good thermoelectric materials[13-18]. For instance, $Ca_5Al_2Sb_6$ has a high thermoelectric performance[19], and especially with its carrier concentration optimized by the partial substitution of Na or Zn on the Ca site, a high figure of ZT>0.6 can be achieved at 1000K[20]. Here, we explore another type of Zintl compounds for their thermoelectric application. These two ternary arsenides $M_2Zn_5As_4$(M=K, Rb) were synthesised[12] last year which adopt a new structure with the space group Cmcm. The M sites are formally 1+ cations while the $(Zn_5As_4)^{-1}$ network is covalently bounded slabs seperated by cationic layers. Band structure calculations on these compounds by the Stuttgart TB-LMTO-ASA program[12] indicate semiconducting behavior with a direct band gap of 0.4 eV which may provide a high Seeback cofficient. Further studies of the detailed bonding and electronic transport may provide insight into the source of the varying thermoelectric performance among the $M_2Zn_5As_4$ compounds.

In this paper, the electric structure and the thermoelectric properties of $M_2Zn_5As_4$(M=K, Rb) are investigated by first-principles method and semiclassical BoltzTraP theory. Different exchange-correlation potentials including the generalized gradient approximation in terms of Perdew-Burke-Ernzerl (PBE-GGA) and the Tran-Blaha modified Becke-Johnson (TB-mBJ) are uesed to determine the band structure of these compounds. The TB-mBJ potential has been shown to provide an accurate account of the electronic structure of correlated compounds using a parameter-free description, also yielding accurate band gaps for most semiconductors[13, 21, 22]. The band structure results for K$_2$Zn$_5$As$_4$ (Rb$_2$Zn$_5$As$_4$) calculated by GGA and TB-mBJLDA show that the band gap are 0.216 (0.189) and 1.094 (0.969) eV, respectively. However, we find that the compounds A$_2$Zn$_5$As$_4$ are semiconductors with indirect band gaps which are different to Ref.12 . The reason might be that some high-symmetry points in irreducible Brillouin zone are not included during the band structure calculation in Ref.12. Also, there are a combination of very narrow bands and more dispersive bands near the valence band edge, which provides a great advantage on the thermoelectric properties of the material. The study of these compounds indicates the thermoelectric parameters including S, thermopower, and $Z_e T = \dfrac{S^2 \sigma T}{\kappa_e}$ are larger in the hole-doping region than the ones in the electron-doping region. Thus, these p-type Zintl phases may be a new type of high thermoelectric materials in the futere applications.

## 2. Computational methods

The experimental structures of M$_2$Zn$_5$As$_4$ (M=K, Rb) are taken as the initial bulk model to relax to find the minimum energy structure which is optimized using the Vienna ab initio simulation package(VASP) based on density functional theory. The exchange-correation energy was calculated by PBE-GGA. Lots of converging tests are performed, and with the cutoff energy of 500 eV and 7×11×7 Monkhorst-Pack special k-points for integrals in the Brillouin zone. The stopping criterion for electronic self-consistent interactions is convergence of the total energy to within 10$^{-6}$ ev. The structure is considered to be in equilibrium when the Hellmann-Feynman forces acting on ions was finally reduced to 0.005 eV Å$^{-1}$. Electronic-structure calculations are performed using the general potential linearized augmented plane-wave (LAPW) method as implemented in the WIEN2K code. Two different exchange-correlation potentials of PBE-GGA and TB-mBJ are used in our calculation. All calculations are fully converged with respect to all the parameters used. In particular, we use $R_{mt}K_{max} = 7.0$, where $R_{mt}$ and $K_{max}$ are the smallest muffin-tin radius and the magnitude of the largest reciprocal-lattice vectors. Self-consistent calculations were done with a *k*- mesh of 10×10×10, and muffin-tin radii of 2.5 a.u. for K, 2.47 a.u. for Zn, and 2.19 a.u. for As. In the calculation for K$_2$Zn$_5$As$_4$, the muffin-tin radii were chosen to be 2.5 a.u. for Rb and Zn, and 2.22 a.u. for As.

The transport properties are obtained from the analysis of band structure results using a semiclassical solution based on Boltzmann theory within the constant scattering time approximation(CSTA) by methods of the BoltzTraP code[23]. The CSTA assumes the relaxation time $\tau$ as energy independent which results in expressions of both the thermopower and the thermoelectric figure of merit with no dependence on $\tau$. This method has been successfully described the transport coefficients of a wide range of thermoelectric materials.

## 3. Results and discussions

The structure type (space group Cmcm, Pearson symbol oC44, Wyckoff sequence g2 f2 e c) of M$_2$Zn$_5$As$_4$ (M=K, Rb) is new. ZnAs$_4$ tetrahedra are linked through corner- and edge-sharing to form a three-dimensional framework, generating large channels extending along the b-direction within which lie the M atoms. The lattice structure and its first Brillouin zone are shown in Fig.1. The optimized equilibrium lattice parameters computed within GGA are determined as a=11.670, b=7.136, c=11.721 for K2Zn5As4 and a=11.759, b=7.196 c=11.854 for Rb$_2$Zn$_5$As$_4$, respectively. These theoretical calculations are slightly larger than the experimental values[12] because the GGA often overestimates the lattice constants.

Fig.2(a) shows the density of states (DOS) of Ca$_5$Al$_2$Sb$_6$. The calculations used by different exchange-correlation potentials show that the shape of DOS is almost identical except for energy shifts. The band gaps obtained from the PBE-GGA and TB-mBJ are 0.225, and 0.521 eV, respectively. In comparison with experimental value 0.5 eV[20], the GGA greatly underestimates the band gap, while the TB-mBJ gives very much improved the band-gap value which is close to the experimental value. Thus, we have reason to believe that TB-mBJ can give a more accurate band gap which plays an important role in the study about thermoelectric properties. As shown in Fig.2(b) and Fig.2(c), the plots indicate the band gap of K$_2$Zn$_5$As$_4$ (Rb$_2$Zn$_5$As$_4$) is 0.216 eV (0.189) with the PBE-GGA and 1.094 eV (0.969) by the TB-mBJ. Unfortunately, as far as we know, there has no experimenal results can be compared with our calculations. Fig.3 is the band structure of $K_2Zn_5As_4$, where the upper plot Fig.3(a) is from Ref.[12] and the lower panel is about our calculation. It can be seen from Fig.3(a) and Fig.3(b), the band structures when the k-path is chosen as Γ-Z-T-Y-Γ-S-R are almost the same with the direct band gap of 0.4 eV, and the valence band maximum (VBM) and the conduction band minimum (CBM) are both at Γ point. However, the result within the GGA demonstrates an indirect band gap from Γ to $F_0$ when more another two high-symmetry points are considered in the first Brillouin zone. The band structures for $K_2Zn_5As_4$ and $Rb_2Zn_5As_4$ computed by TB-mBJ are shown in Fig.4, where the dot line represents the Fermi level. Comparing Fig.3(b) and Fig.4(a) about the band structure of $K_2Zn_5As_4$, the same spectroscopies are acquired with the various functionals, the size of the gap being the main difference. The valence-bands near the VBM are made up by three bands. Two small dispersive bands of them degenerated at Γ point form heavier bands, which should lead to large thermopower. Another strong dispersive band is approximately parabolic near Γ point become a light band which can be easily accessed for hole transport to promote carrier mobility. Thus, the combinnation of heavy and light bands in valence bands is favorable to the high thermoelectric performance which has been previously discussed[13, 21].

Due to the similar electronic structure of $K_2Zn_5As_4$ and $Rb_2Zn_5As_4$ in Fig.4, we take $Rb_2Zn_5As_4$ as an example to introduce the characteristics. The calculated projected DOS is shown in Fig.5 with the curve revealing nearly small contribution Rb-based states. In the valence band, the peak of Rb-p states is located at -11.46 eV, where a sharply increased contribution of As-s states is observed between -11.46 and -9.96 eV. The Zn-d states are found as narrow bands in the range of -7.402 to -6.51 eV. Energies much below $E_f$ ensure that these bands almost retain anatomic character. The edges of both the valence and conduction bands are derived from hybridized As and Zn orbitals which means the forming of covalent As-Zn bonds. The Rb-s bands are found in the conduction bands implying the atom is essentially fully ionic. The coexist of covalent band and ionic band is consistent with the bonding picture for the Zintl phase. One can also see from the DOS curve that there is an rapidly increase away from the band edge for both the valence and conduction bands. This means that the material doped by hole or electron can exhibit a high thermoelectric performance. Thermoelectric properties of the p-doping compound may be better due to a more sharp peaks in valence bands between -2.24 eV to $E_f$. The detailed expression about the thermoelectricity is in the following section.

The transport properties of $Rb_2Zn_5As_4$ based on the calculated TB-mBJ electronic structure are calculated by the semiclassical Boltzmann theory. Yet, the value of the thermoelectric figure of merit $ZT = \frac{S^2 \sigma T}{\kappa}$ is independent of the scattering time chosen and so are the thermopower and $\sigma/k$, as long as the CSTA is retained. The lattice component of thermal conducitivity $\kappa_l$ caused by the phonon scattering is not considered in our calculations. As an estimate of $Rb_2Zn_5As_4$ efficiency, the thermoelectric figure of merit can be expressed as $Z_e T = \frac{S^2 \sigma T}{\kappa_e}$, which are called " maximum" thermoelectric figures of merit. We also used the rigid-band approach to research how thermoelectric performance depended on the doping level. That is to say the thermoelectric properties are calculated using the undoped band structure with appropriate temperature dependent shifts of the Fermi level to correspond to the doping level. In Fig.6, we present the thermoelectrical cofficients $S$, $S^2\sigma/\tau$, the carrier concentration $n$ and $Z_e T$ as a function of the chemical potential calculated at different temperature, where two dot lines represent the VBM and CBM, respectively. In order to discuss the thermoelectric bahavior, we write the expressions for the seeback cofficient $S$, and electronic conductivity $\sigma$ as follows.

$$S = \frac{8\pi^2 k_B^2}{3eh^2} m^* T (\frac{\pi}{3n})^{\frac{2}{3}} \qquad (2)$$

$$\sigma = ne\mu \qquad (3)$$

Where $k_B$ is the Boltzmann constant, $e$ is the electronic charge, $h$ is the Planck constant, $n$ is carrier concentration, $m^*$ is the carrier's effective mass, and $\mu$ is carrier mobility. As shown in Fig.6(a), it shows that the value of $S$ in the hole-doping region is much bigger than the one in the e-doping region. From equation (2), we can see that the larger $S$ near the VBM is a result of the heavy effective mass $m^*$ which is related to small disperasive bands provided mainly by the As-p orbital. As a result of the increase of the hole-doping, the $S$ decreases sharply and the sign of S changes from positive at low doping to negative at higher doping. A similar feature happens in the electron-doped region: S goes from negative to positive when the system is further electron doped(about 0.4 eV above the bottom of the conduction band). The sign of change has been observed experimentally for $La_2NiO_{4+\delta}$ when the value δ～0[24]which has been proved theoretically[25]. This indicates the electronic structure will be rearranged by a heavy doping which leads to the observed complex thermal dependence of the Seecback coefficient. Besides, we note an interesting phenomenon that the $S$ drops with the increasing temperature when the chemical potential $E > -0.059\ eV$. However, the higher temperature corresponds to the higher S in the case of $E < -0.059\ eV$. The reason is that the difference of the carrier concentrations is smaller with the increasing of the hole-doping level for different temperatures which can be seen from Fig.6(b). The inset is the blowup of the carrier concentrations. Thus, on the basis of the equation (2) the influence of the temperature on the $S$ plays a important role when the material is further hole-doped(ie. $E < -0.059\ eV$). Due to the increase of the carrier concentration, the value of $S^2\sigma/\tau$ is larger at a higher temperture(shown in Fig.6(c)). It is also observed that $S^2\sigma/\tau$ for p-type doping is larger than that of n-type doping due to the combination of the heavy bands and light bands near the VBM. In the hole-doping region, the $S^2\sigma/\tau$ increases firstly and reaches a maximum at $E = -0.158\ eV$ for T=800 K. Then, it descreases rapidly and becomes zero at about $E = -0.567\ eV$ when the $S$ sign changes from positive to negative. Similar results for T=300, 400, and 600 K are presented by different color lines. The high value of thermopower with respect to relaxation time is comparable to conventional high thermoelectric p-type materials PbTe[2]. Furthermore, these materials are expected to have low thermal conductivity due to their complex lattice picture. For this reason, the compound may represent a useful example for finding new thermoelectric materials. Fig.6(d) shows reasonably large value of the $Z_eT$ for Rb$_2$Zn$_5$As$_4$. However, the $Z_eT$ value will be smaller when the thermal conductivity due to phonons part is included. In any case, the results will be significant when the electron thermal conductivity $\kappa_{el}$ is comparable to the lattice thermal conductivity $\kappa_l$. The $Z_eT$ value for p-type doping is larger than that of n-doped doping, and the corresponding maximum value $Z_eT$ is 0.92 and 0.75, respectively. The reason is, as described above, the combination of heavy and light bands in valence bands since the strong dispersive band (light bands) favors reasonable $\sigma$, while the weakly dispersive band (heavy bands) sets a low-energy scale that can enhance the *S*. From Fig.6(b) and Fig.6(d), we can see that the carrier concentrations corresponding to the large $Z_eT$ are in the range of n~(1-4)×10$^{20}$cm$^{-3}$. Such a concentration range falls just between 10$^{19}$ and 10$^{21}$ carriers per cm$^{-3}$ for good thermoelectric materials. Consequently, it is more conducive to improve the figure of the merit when the material is doped by the hole, and the high thermoelectric performance of M$_2$Zn$_5$As$_4$(M=K, Rb) would be attractive for its potential applications.

## 4. Summary

Our ab initio calculations for these compounds indicate some promising features of hole-doped M$_2$Zn$_5$As$_4$ as a possible thermoelectric materials. Due to the combination of very narrow bands and more dispersive bands near the valence band edge, the high thermopower are expected. The thermoelectric transport properties study of these compounds indicates the values of $Z_eT$ are larger in hole-doping region than the value in electron-doping region. With the increasing hole-doped level, the $Z_eT$ value increases slowly and up to a maximum, then it descrease droply and becomes zero simultaneously with the *S* sign changes from postive to negative. Simlar results can be found in the electron-doping region. Thus, these compounds would be suitable high performance thermoelectric materials at a low hole-doping level.

Acknowledgments

This work was supported partly by the National Natural Science of China (Grant No.11047108, 11147197, 11005003), partly by the Research Project of Basic and Cutting-edge Technology of Henan Province, (Grant No. 112300410183), and partly by the Education Department and Department of Henan Province (Grant No.2011B140002 ).

Figure captions

Figure 1. (Color online) (a) Viewed down the b-direction of the Structure of $M_2Zn_5As_4$ (M= K, Rb). The red circles are M atoms, the purple circles are As atoms, and the grey circles are As atoms. (b) The first Brillouin zone of $M_2Zn_5As_4$.

Figure 2. Comparation of the PBE-GGA and TB-mBJ electronic density of states of $Ca_5Al_2Sb_6$, $K_2Zn_5As_4$, and $Rb_2Zn_5As_4$.

Figure 3. Comparation of the band structure in Ref.[12](a) with our calculation (b).

Figure 4. The band structure of (a) $K_2Zn_5As_4$ (b) $Rb_2Zn_5As_4$ calculated by TB-mBJ.

Figure 5. (Color online) Calculated projected DOS for $Rb_2Zn_5As_4$.

Figure 6. Thermoelectric transport coefficients of $Rb_2Zn_5As_4$ as a function of chemical potential. (a) Seebeck coefficients $S$, (b) Carriers concentration $n$ (c) Thermopower with respect to relaxation time $S^2\sigma/\tau$. (d) the big " maximum" thermoelectric figures of merit $Z_eT$.

Fig.1

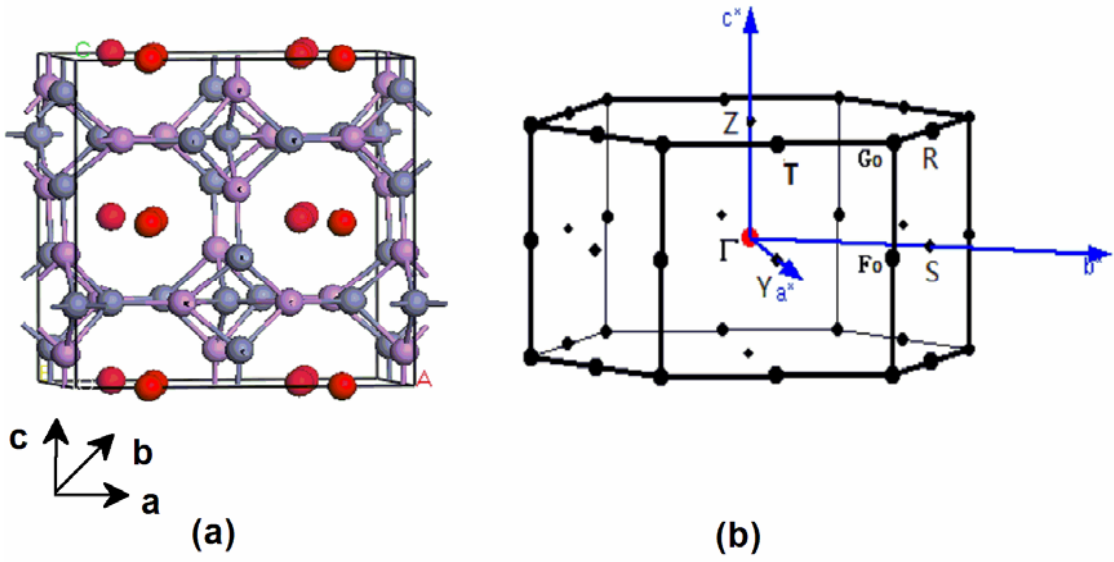

(a)  (b)

Fig.2

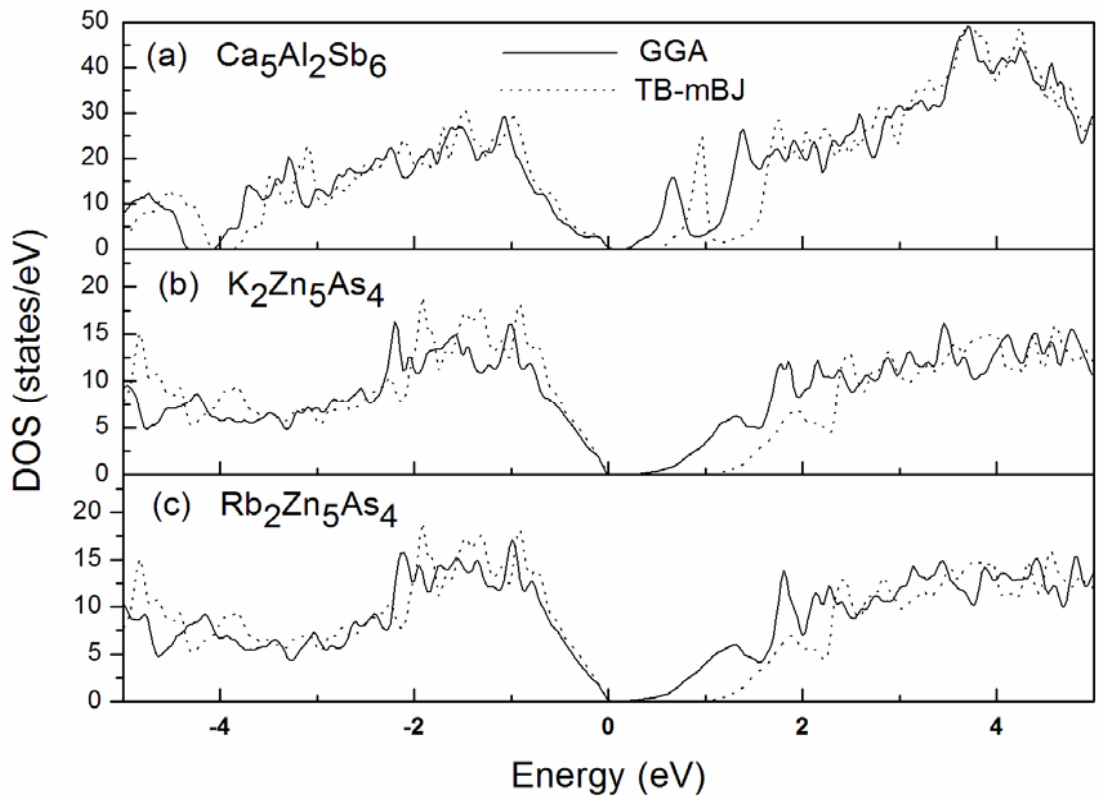

Fig.3

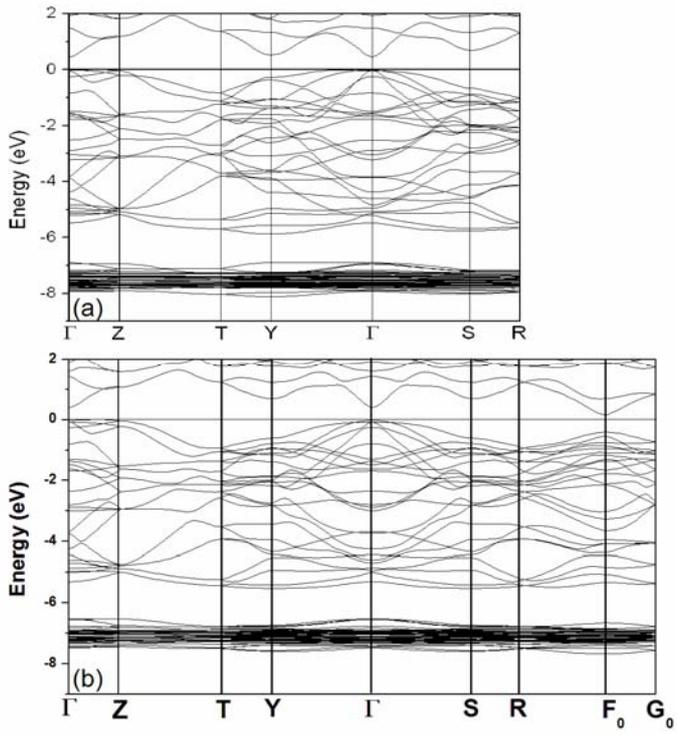

Fig.4

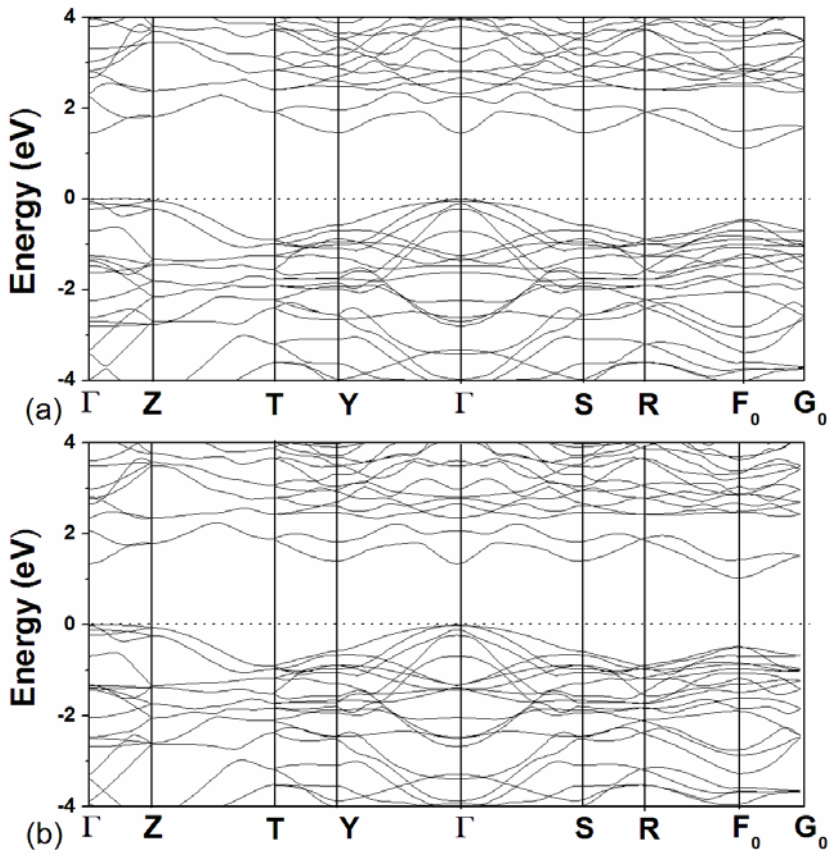

Fig.5

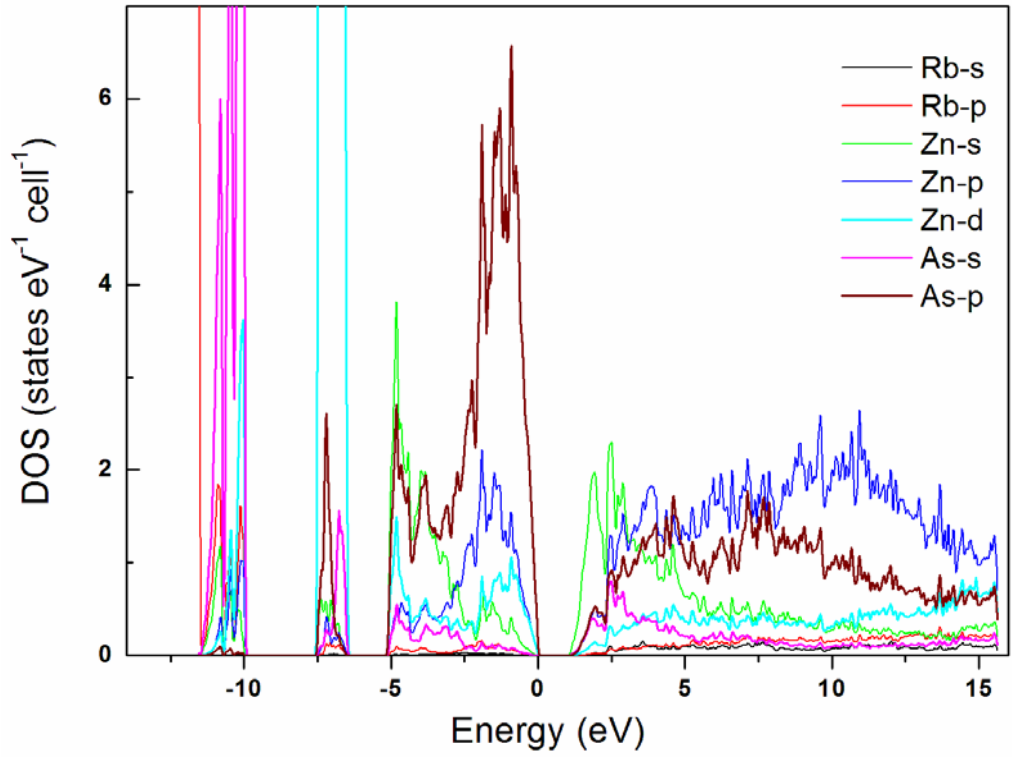

Fig.6

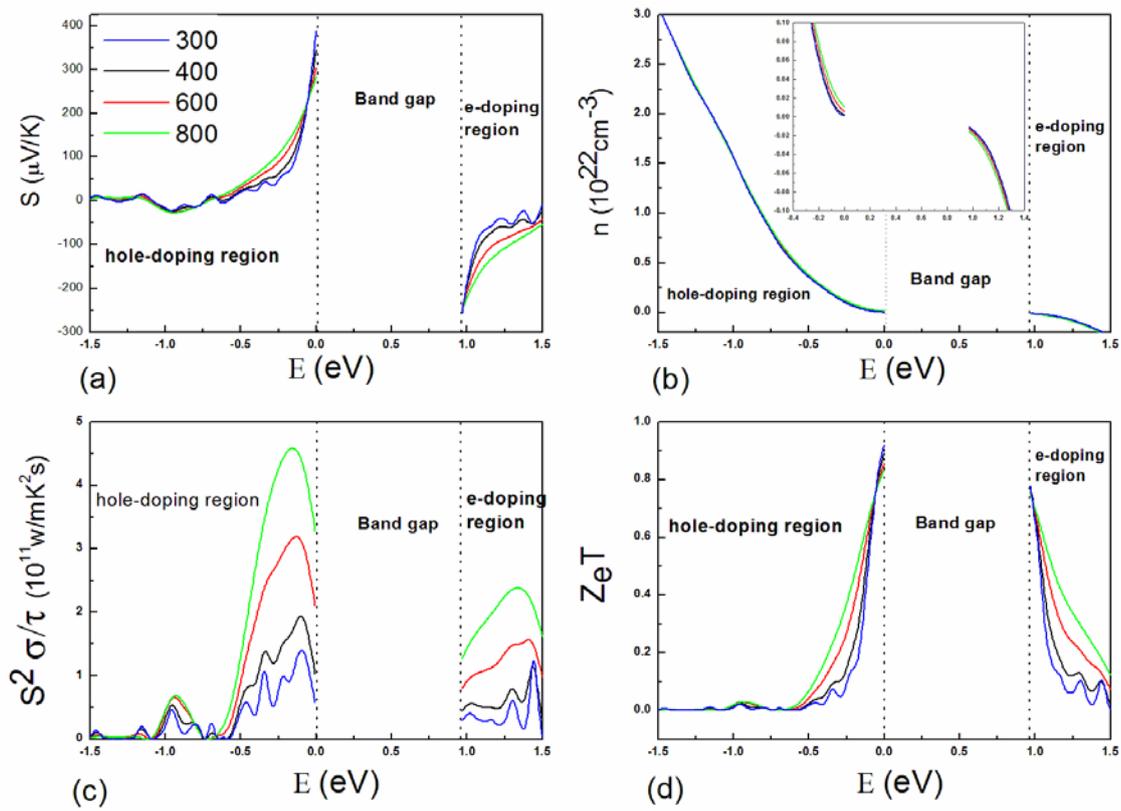